\documentclass[%
 reprint,
 amsmath,amssymb,
 aps,
]{revtex4-1}

\usepackage{graphicx}
\usepackage{dcolumn}
\usepackage{bm}

\begin{document}


\title{Trivial Entropy of Matter in Gravitation}

\author{Jun Chen}
\affiliation{Zhejiang University, Hangzhou, China}
\email{junc@zju.edu.cn}

\date{\today}

\begin{abstract}
Expanding the black hole thermodynamics from the horizon to achronal Cauchy hypersurface, the general relation between the Einstein equation and thermodynamics is established. Starting from trivial entropy that is generalized by Bekenstein-Hawking entropy, the entropic mass of matter emerges naturally together with Unruh temperature. The key idea is that the cause of mass formation comes down to trivial entropy, and mass density is just the external manifestation of mass. The full Einstein equation with the cosmological constant is derived from the requirement that entropic mass and proper mass are equivalent. This perspective suggests that trivial entropy that causes mass in gravitation may be the best choice for the origin of space-time geometry.
\end{abstract}

\pacs{Valid PACS appear here}
\maketitle


\section{Introduction}
\label{sec:intro}

As a basic force, nature gravity has been widely discussed by many physicists because of its intrinsic. So far, the most successful example of describing gravity is the Einstein equation~\cite{b} that was originally derived by Einstein from the local conservation of the stress tensor and the contracted Bianchi identity. Several years later, in 1995, Jacobson~\cite{d} observed that the Einstein equation relates to thermodynamics near causal Rindler horizons. Consider that the equilibrium thermodynamic relation holds for all the local Rindler causal horizons in each spacetime point. In thermodynamics of spacetime, the Einstein equation considered as an equation of state is successfully derived by introducing Unruh temperature~\cite{c} that is demanded to see by an observer with acceleration inside the horizon.

By 2011, Verlinde~\cite{a} proposed to extend the relation between thermodynamics and the Einstein equation to holographic screens that bound emergent spacetime, even far from a horizon. By drawing on the holographic principle supporting by the AdS/CFT correspondence, Verlinde believes that the essence of gravity is entropic force. With the assumption of a linear relation between the displacement of the holographic screen and entropy, together with the equipartition rule, Newton's law of gravitation and the second law of Newton can be derived by entropic force. Of course, as well as the Einstein equation. Although Verlinde extended the logic of Jacobson to far away from causal Rindler horizons, it inevitably introduced the assumption of holographic theory that associates information bits describing entropy with the degrees of freedom on holographic screen. This leads us to rethink the more general connection between thermodynamics and the Einstein equation, without the bounds of horizon, nor the assumption of holographic theory.

In this paper, we propose to further extend the logic of Verlinde to spacelike Cauchy hypersurfaces $\Sigma$. In 1965, the terminology of Cauchy hypersurface was first introduced by Penrose~\cite{h} in general relativity. For a spacetime $(M,g_{ab})$, the achronal closed set $\Sigma$ contained in $M$ is called Cauchy hypersurface if $D(\Sigma)=M$. Since $\Sigma$ is achronal, it may represents an ``instant of time" throughout the universe, which will help us build a general relation between thermodynamics and the Einstein equation.

Abandoning the constraints of horizons or holographic screens, we manage to redefine the Bekenstein-Hawking entropy $S_{BH}$~\cite{e,f} and Unruh temperature $T_U$ from the horizons to the achronal Cauchy hypersurfaces. There is no doubt that the acceleration carried by Unruh temperature measures by an observer at a distance away from the Cauchy surfaces. In addition, the horizon area $A_H$ of Bekenstein-Hawking entropy will be replaced with a topological 2-sphere area $\mathcal{A}_C$, where the topological 2-sphere is the boundary of the Cauchy hypersurface. Under the condition that entropy is proportional to area including the factor $1/4$, we naturally generalize Bekenstein-Hawking entropy as trivial entropy $S_{TR}$.

So far, with the Cauchy surface as the starting point, we have established a series of thermodynamic quantities. Since we try to get a general relation between thermodynamics and the Einstein equation, then gravitational field equation is a good tool to test our conjecture. If we can derive the Einstein field equations from thermodynamics near achronal Cauchy surfaces, then it shows that trivial entropy is indeed well-defined and universally applicable.

\section{Thermodynamics near achronal Cauchy surfaces}
\subsection{Entropy and temperature}

The entropy is a central notion in our discussion, this basic concept can be elaborated by thermodynamics in a homogeneous system. As a state function, the entropy $S(E,V)$ is associated with the total system energy $E$ and the volume $V$. In this case, one can give $S(E,V) = k_B \operatorname{ln} \Omega(E,V)$ that is used to describe the degree of order of the system, where $k_B$ is Boltzman's constant and $\Omega(E,V)$ denotes the thermodynamic probability. If one agrees with this definition, then a corresponding $\Omega(E,V)$ for a macroscopic state always can be found. 

The first law of thermodynamics yields tells us $\delta Q = dE + p dV$ in the enclosed system, where $Q$ is the total heat and $p$ is the system pressure. By introducing the equilibrium thermodynamic relation $\delta Q = T dS$ connecting temperature $T$, heat $Q$, and the change of entropy $dS$, one can deduce the relation $dE = T dS$ in the isochoric process ($dV=0$). On the other hand, energy can be expressed in the form of external force $F$ doing work as $dE = F dx$. By combining the above two relations, one can give equation in 1-dimensional case~\cite{a}
\begin{equation}
F dx = T dS
\label{entropy}
\end{equation}

It remains to maintain the system temperature by introducing the heat flow. From the perspective of an observer with the constant acceleration who just hovers near Cauchy surface, the same vacuum fluctuations with a thermal character will exert the Unruh effect~\cite{c}. For observers with different accelerations, the measured temperatures are different. And this temperature is in good agreement with the expected results of immersing the particle detectors in a thermal bath of scalar photons. For consistency, we consider the temperature as the Unruh temperature associated with such a uniformly accelerated observer
\begin{equation}
T = T_U = \frac{\hbar}{2 \pi c k_B} a
\label{temperature1}
\end{equation}
A non-zero acceleration can effectively ensure that the Unruh temperature is non-zero, where the temperature $T$ at this time is caused by an acceleration $a$. Due to the emergence of acceleration, we better substitute the acceleration in the 4-dimensional space-time continuum, which is ``absolute'' compared to that in the 3-dimensional space.

\subsection{The Newtonian potential of space-time}

For the following discussion, from now on, we will use tensor notation.
Now, consider a static and asymptotically flat spacetime background, the fact it is natural to define the acceleration through Newton's potential. As Wald~\cite{b}, the Newton's potential is
\begin{equation}
\phi = \frac{1}{2} \operatorname{ln}(-\xi^a \xi_a)
\end{equation}
where its exponent $e^{\phi}$ represents a redshift factor and $\xi^a$ is a global time-like Killing vector field. In a neighborhood of infinity, $\xi^a$ is normalized so that the redshift factor approaches $1$ at infinity.

The 3-acceleration of the particle can be defined in terms of the gradient of the Newton's potential $\phi$ as
\begin{equation}
a^b \equiv -\nabla^b \phi = -e^{-\phi} \nabla^b e^{\phi}
\end{equation}

For an observer of an arbitrary steady-state frame of reference, its world line coincides with the integral curve of the Killing field, and its 4-velocity meets the condition $U^b = e^{-\phi} \xi^b$ based on the relation $U^b U_b = -1$. Hence, the 4-acceleration of the particle follows an orbit of the Killing vector field $\xi^a$, which can be expressed by the 4-velocity $U^b$ of the particle. The 4-acceleration is given by
\begin{equation}
A^b \equiv U^a \nabla_a U^b = e^{-2\phi} \xi^a \nabla_a \xi^b
\label{acceleration1}
\end{equation}
We can further rewrite the last equation (\ref{acceleration1}) by using Killing's equation $\nabla_a \xi_b = \nabla_{[a} \xi_{b]}$
\begin{equation}
\begin{aligned}
A^b &= -e^{-2\phi} \xi^a \nabla^b \xi_a = -\frac{1}{2} e^{-2\phi} \nabla^b (\xi^a \xi_a) \\
&= -\frac{1}{2} e^{-2\phi} \nabla^b (-e^{2\phi}) = e^{-\phi} \nabla^b e^{\phi} \\
&= \nabla^b \phi
\end{aligned}
\label{acceleration2}
\end{equation}
The difference between 3-acceleration and 4-acceleration is a minus sign, which indicates that the expression of the 4-acceleration is well-defined.

\subsection{Trivial entropy}

With the defined 4-acceleration (\ref{acceleration1}), the equation (\ref{entropy}) is generalized to
\begin{equation}
T \nabla_b S = A_b \equiv e^{-2\phi} \xi^a \nabla_a \xi_b.
\label{assume}
\end{equation}
On the right-hand side of this equation, according to Newton's law of inertia $F_b=m A_b$, the disappearing mass $m$ is interpreted as the force that is exerted to hold a unit test mass in place. The notion of ``stay in place'' means that a unit test mass follows an orbit of the Killing field $\xi^a$. On the left-hand side of this equation, the gradient of the entropy is $\nabla_b S$ for an isolated system with temperature $T$. In this case, the formula gives a physical picture, which connects the entropy and the mass universally.

So far we have argued that the connection between mass and entropy is universal in the static and asymptotically flat space-time. Now recall where this kind of relation has still ever appeared. In strong gravitational fields, near black holes, this relation also exists, but that relates between Bekenstein-Hawking entropy~\cite{f} and the mass of black holes. The Bekenstein-Hawking formula of black hole gives
\begin{equation}
S_{BH} = \frac{c^3 k_B}{4 \hbar G} A_H
\end{equation}
where $A_H$ denotes the area of the event horizon. In other words, black hole entropy can be expressed in terms of the horizon surrounding the black hole. For a Schwarzschild-type black hole, the relation between its mass and the radius of horizon is given by formula $R = 2 G M / c^2$. So at least for the case of black holes, the connection between mass and entropy is already obvious.

From the equation (\ref{assume}), this connection seems to be trivial and reasonable, not limited to the case of black holes. We assume that black hole entropy can be generalized to trivial entropy that is applicable to any matter. First, all the mass of a black hole is surrounded by the area of its horizon, so the black hole entropy element $dS_{BH}$ is proportional to the area element $dA_H$. However, other objects do not have such a horizon, and what can they be surrounded by? A topological 2-sphere $\mathcal{S}$ as the boundary of Cauchy hypersurface $\Sigma$, which encloses all the sources. Naturally, we give the trivial entropy of matter in the differential form
\begin{equation}
dS_{TR} = \frac{c^3 k_B}{4 \hbar G} d\mathcal{A}_C
\label{trivial}
\end{equation}
where $d\mathcal{A}_C$ is the area element on a topological 2-sphere $\mathcal{S}$. This form of trivial entropy is able to link the entropy and mass, which will be the core of our argument.

\begin{figure}
	\includegraphics[scale=0.5]{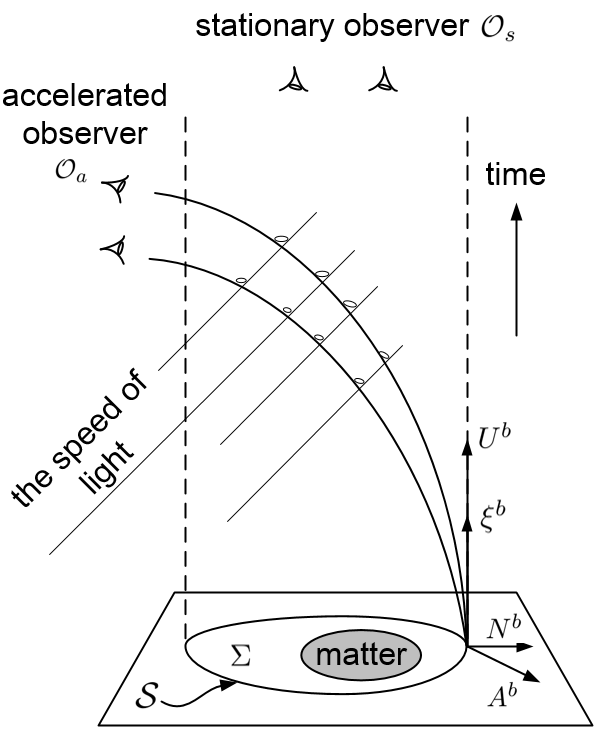}
	\caption{Entropic mass $M_e$ is defined as the integral on any topological 2-sphere $\mathcal{S}$ surrounding all the matter. Cauchy hypersurface $\Sigma$ as the boundary of $\mathcal{S}$ is always perpendicular to the 4-velocity $U^b$ of a stationary observer $\mathcal{O}_s$. A uniformly accelerated observer $\mathcal{O}_a$, whose world line is tangent to that of $\mathcal{O}_s$ at $\mathcal{S}$. The temperature $T$ and trivial entropy $S_{TR}$ both are measured on $\mathcal{S}$ by $\mathcal{O}_a$.}
	\label{cauchy}
\end{figure}

\section{Space-time continuum and identity of mass}
\subsection{Proper mass}

So far, with the well-defined relations above, we can discuss the multipole expansion definition of mass and give a physical picture. Here, we mainly consider the duality of mass caused by internal and external properties respectively. According to the definition of trivial entropy we generalized in Section 2.3, mass can be expressed in the form of entropy and temperature. We attribute entropy to the internal property of mass that is called as entropic mass $M_e$. Hence, the internal property of mass is a kind of thermodynamic attribution. Corresponding to the internal property is the external property, one refers to the external property of mass as proper mass $M_p$ that is broadly defined. Given the mass-energy density of matter $\rho$, $M_p$ can be expressed by the integral relation $\int \rho dV$ over the 3-dimensional space $V$.

Now we introduce a stress-energy tensor $T_{ab}$. In this way, the mass-energy density $\rho$ can be defined as $\rho = T_{ab} n^a \xi^b$ associated with a time-translation Killing field $\xi^b$. By combining mass-energy equation, the total proper mass equals to
\begin{equation}
M_p = \frac{1}{c^2} E \equiv \frac{1}{c^2} \int_{\Sigma} T_{ab} n^a \xi^b d\mathcal{V}
\label{mp}
\end{equation}
where $\Sigma$ is a space-like Cauchy hypersurface and $n^a$ denotes the unit future pointing normal to $\Sigma$.

Global conservation $\nabla^a \left(T_{ab} \xi^b\right) = \left(\nabla^a T_{ab}\right) \xi^b + T_{ab} \nabla^a \xi^b = 0$ of energy-momentum of matter implies that $T_{ab} \xi^b$ is divergence free. So $\int_{\Sigma} T_{ab} \xi^b n^a d\mathcal{V}$ is conserved, i.e., independent of choice of Cauchy surface $\Sigma$. Therefore, any Cauchy surface $\Sigma$ is valid for the above formula.

\subsection{Entropic mass}

For the derivation of entropic mass, we need to use the well-defined relations from Section 2 again. The overall background we consider is always a static and asymptotically flat spacetime, in other words, spacetime tends to be flat at infinity.  Consider a topological 2-sphere $\mathcal{S}$ lying in Cauchy hypersurface orthogonal to time-like Killing field $\xi^a$, and this 2-sphere encloses the total proper mass $M_p$ as the boundary of a space-like hypersurface $\Sigma$.

In space-time continuum, the Unruh law (\ref{temperature1}) now is re-expressed in tensor form
\begin{equation}
T = \frac{\hbar}{2 \pi c k_B} e^{\phi} N^b \left(e^{-2\phi} \xi^a \nabla_a \xi_b\right)
\label{temperature2}
\end{equation}
where the acceleration $a$ should be replaced by 4-acceleration $\nabla_b \phi$ or $e^{-2\phi} \xi^a \nabla_a \xi_b$ based on the relations (\ref{acceleration1}) and (\ref{acceleration2}), and $N^b$ is the unit ``outward pointing'' normal to $\mathcal{S}$. In addition, we insert a redshift factor $e^{\phi}$, because the temperature differs from the local temperature by a factor of $e^{\phi}$ if we choose to deduce the temperature which must be applied by a distant observer at infinity.

The basic principle that still needs to play a role in our derivation is: the thermodynamic relation $\delta Q = T dS$, whose integral form $\int_{\mathcal{S}} T dS_{TR}$ is interpreted as the total entropic mass that is shown in FIG~\ref{cauchy}. For an isolated system containing matter, the source of all its heat $Q$ is the mass, i.e., entropic mass, and it is easily seen to agree with the definition based on the argument in Section 2.3. By combining the relation $Q = M_e$ together with (\ref{temperature2}) and (\ref{trivial}), we can deduce an expression of satisfying entropic mass
\begin{equation}
\begin{aligned}
M_e &= \int_{\mathcal{S}} T dS_{TR} \\
&= \int_{\mathcal{S}} \frac{\hbar}{2 \pi c k_B} e^{\phi} N^b \left(e^{-2\phi} \xi^a \nabla_a \xi_b\right) \frac{c^3 k_B}{4 \hbar G} d\mathcal{A}_C \\
&= \frac{c^2}{8\pi G} \int_{\mathcal{S}} e^{-\phi} N^b \xi^a \nabla_a \xi_b  d \mathcal{A}_C
\end{aligned}
\label{mass}
\end{equation}
With the help of the definition of 4-acceleration with the Newtonian potential, the last equation can also be written to
\begin{equation}
M_e = \frac{c^2}{8\pi G} \int_{\mathcal{S}} N^b \nabla_b \phi  d \mathcal{A}_C
\end{equation}
In the exterior and vacuum region, the Newtonian potential $\phi$ satisfies Laplace's equation $\nabla^2 \phi = 0$, so the integral is independent of choice of 2-sphere $\mathcal{S}$, which also implies that the multipole expansion definition of mass is reasonable.

We can further express the equation (\ref{mass}) as the form of Komar mass $M_k$ that is first given by Komar~\cite{g}, which provides a fully satisfactory notion of the total mass in stationary and asymptotically flat space-time.
\begin{equation}
\begin{aligned}
M_e &= \frac{c^2}{8\pi G} \int_{\mathcal{S}} e^{-\phi} N^d \xi^c \nabla_c \xi_d  \epsilon_{ab} \\
&= \frac{c^2}{8\pi G} \int_{\mathcal{S}} \epsilon_{ab} N_d u_c \nabla^c \xi^d \\
&= -\frac{c^2}{16\pi G} \int_{\mathcal{S}} \epsilon_{abcd} \nabla^c \xi^d  \\
&\equiv\frac{1}{2} M_k
\label{mass1}
\end{aligned}
\end{equation}
Where in the first step $\epsilon_{ab}$ is the volume element represented by $d \mathcal{A}_C$ on the 2-dimensional submanifold $\mathcal{S}$, in the second step 4-velocity $U^b = e^{-\phi} \xi^b$ is used, and the detailed derivation of the third step is seen as follows. Let $\alpha_{ab} \equiv \epsilon_{abcd} \nabla^c \xi^d$ where $\epsilon_{abcd}$ is the volume element on 4-dimensional space-time associated with the spacetime metric, then the restriction of $\alpha_{ab}$ on $\mathcal{S}$ is viewed as 2-form field $\tilde{\alpha}_{ab}$ which differs from $\epsilon_{ab}$ by a function factor $f$, i.e., $\tilde{\alpha}_{ab} = f \epsilon_{ab}$. Contraction of this identity using $\epsilon^{ab} = N_e U_f \epsilon^{feab}$ leads the right-hand side of this identity to give $\epsilon^{ab} f \epsilon_{ab} = 2f$, and the left-hand side of this identity gives
\begin{equation}
\begin{aligned}
\epsilon^{ab} \tilde{\alpha}_{ab} &= \epsilon^{ab} \alpha_{ab} \\
&= N_e U_f \epsilon^{feab} \epsilon_{abcd} \nabla^c \xi^d \\
&= -4 N_e U_f \delta^f_{\ [c} \delta^e_{\ d]} \nabla^c \xi^d \\
&= -4 N_e U_f \delta^f_{\ c} \delta^e_{\ d} \nabla^c \xi^d \\
&= -4 N_d U_c \nabla^c \xi^d.
\end{aligned}
\end{equation}
Here in the fourth step Killing's equation $\nabla^c \xi^d = \nabla^{[c} \xi^{d]}$ is used. Therefore, we obtain $f = -2 N_d U_c \nabla^c \xi^d$ and complete the above proof.

\subsection{Emergence of the Einstein equations}

In order to show the rationality and universality of trivial entropy, it is important to realize that the full Einstein equation can be deduced by the equivalence of proper mass and entropic mass. Above, we mentioned the duality of mass, which outward manifests as proper mass (\ref{mp}) defined by mass-energy density of matter and inward manifests as entropic mass (\ref{mass1}) defined by trivial entropy and Unruh temperature. Although the forms of mass are different, the results of different representations are the same, i.e., here the equivalence of proper mass and entropic mass, due to the principle of identity of mass.

With the help of $\bm{\alpha} = \alpha_{ab} \equiv \epsilon_{abcd} \nabla^c \xi^d$, the equation (\ref{mass1}) will be further deduced as
\begin{equation}
\begin{aligned}
M_e &= -\frac{c^2}{16\pi G} \int_{\mathcal{S}} \bm{\alpha} \\
&= -\frac{c^2}{16\pi G} \int_{\Sigma} d\bm{\alpha} \\
&= -\frac{3 c^2}{16\pi G} \int_{\Sigma} \nabla_{[e} \left( \epsilon_{ab]cd} \nabla^c \xi^d \right)\\
&= -\frac{c^2}{8\pi G} \int_{\Sigma} R^d_{\ f} \xi^f \epsilon_{deab} \\
&= \frac{c^2}{8\pi G} \int_{\Sigma} R_{ab} n^a \xi^b d\mathcal{V}
\label{mass2}
\end{aligned}
\end{equation}
Where in the second step Stokes's theorem is applied to convert to a volume integral over $\Sigma$ because $\Sigma \cup \mathcal{S}$ is a compact manifold with boundary, in the third step the definition of outer differential form is used, in the fifth step $\epsilon_{abc} = n^d \epsilon_{dabc}$ is the natural volume element on $\Sigma$ represented by $ d\mathcal{V}$ ($n^a$ is the unit future pointing normal to $\Sigma$), and the detailed derivation of the fourth step is seen as follows. Contraction of $\nabla_{[f} \left( \epsilon_{ab]cd} \nabla^c \xi^d \right)$ using $\epsilon^{efab}$ leads to
\begin{equation}
\begin{aligned}
\epsilon^{efab} \nabla_{[f} \left( \epsilon_{ab]cd} \nabla^c \xi^d \right) &= \epsilon^{efab} \epsilon_{abcd} \nabla_f \nabla^c \xi^d \\
&= -4 \nabla_f \nabla^{[e} \xi^{f]} \\
&= 4 \nabla_f \nabla^f \xi^e \\
&= -4 R^e_{\ f} \xi^f
\end{aligned}
\label{mass3}
\end{equation}
Here in the third step Killing's equation $\nabla^f \xi^e = -\nabla^{[e} \xi^{f]}$ is used and in the fourth step the relation $\nabla^a \nabla_a \xi^b = -R^b_{\ a} \xi^a$ is used where $R_{ab}$ is Ricci tensor. Then, by multiplying equation (\ref{mass3}) by $\epsilon_{elmn}$ and contracting over $e$, the right-hand side of this equation gives $-4 R^e_{\ f} \xi^f \epsilon_{elmn}$ and the left-hand side of this equation gives $-6 \nabla_{[l} \left( \epsilon_{mn]cd} \nabla^c \xi^d \right)$. Hence, we finish the above proof.

By being equivalent to proper mass (\ref{mp}) and entropic mass (\ref{mass2}), we lead to the relation
\begin{equation}
\frac{8\pi G}{c^4} \int_{\Sigma} T_{ab} n^a \xi^b d \mathcal{V} = \int_{\Sigma} R_{ab} n^a \xi^b d \mathcal{V}
\label{key}
\end{equation}
As any Cauchy surface $\Sigma$ is valid for the above formula, now the relation (\ref{key}) is only valid if $(8\pi G/c^4) T_{ab} n^a \xi^b = R_{ab} n^a \xi^b$ for all time-like $\xi^b$ and $n^a$, which implies that $(8\pi G/c^4) T_{ab} = R_{ab} + f' g_{ab}$ for a function $f'$. Local conservation $\nabla^a T_{ab} = 0$ of energy-momentum of matter implies that $T_{ab}$ is divergence free. So the right-hand side gives $\nabla^a R_{ab} + \nabla^a \left(f' g_{ab}\right)=0$. By combining Bianchi identity $\nabla_{[a} R_{bc]d}^{\ \ \ \ e}=0$ and the relation $R_{abc}^{\ \ \ \ d}=-R_{bac}^{\ \ \ \ d}$ and contracting over $a$ and $e$ where $R_{abc}^{\ \ \ \ d}$ is Riemann tensor, one obtains $\nabla_a R_{bcd}^{\ \ \ \ a}+\nabla_b R_{cd}-\nabla_c R_{bd}=0$. After raising the index $d$ with the metric and contracting over $b$ and $d$, we find $\nabla^a R_{ab} -  g_{ab} \nabla^a R /2=0$ and $f'=-R/2+\Lambda$ for a constant $\Lambda$. Thus, the full Einstein equation with the cosmological constant emerges:
\begin{equation}
\frac{8\pi G}{c^4} T_{ab} = R_{ab} - \frac{1}{2} R g_{ab} + \Lambda g_{ab}
\end{equation}

\section{Conclusion}

Starting from the fundamental relation $\delta Q = T dS$ on achronal Cauchy surface, we see a general connection between entropy and mass, which gives us reason to believe that the mass of all the matter can be represented by entropy. This entropy is called as trivial entropy that is generalized from the black hole entropy, which implies that black hole entropy applies not only to black holes, but also to all matter. Taking this as a foothold, we derive the entropic mass that is represented by trivial entropy and Unruh temperature. We attribute the representation of entropy to the internal property of mass, and in addition to the representation of mass density to the external property of mass. Unifying these two properties, we naturally derive the full Einstein equation with the cosmological constant. The derivation of the field equation will bring two benefits: on the one hand, this shows that trivial entropy and entropic mass are well-defined and universal in gravitation; on the other hand, the origin of space-time curvature in field equation seems to be trivial entropy.

\end{document}